\documentclass[11pt,a4paper]{article}
\usepackage{graphicx}
\begin{document}
%
%%%%%%%%%%%%%%%%%%%%%%%%%%%%%%%%%%%%%%%%%%%%%%%%%%%%%%%%%%%%%%%%%%%%%%
%
\title{\bf Analytic queueing model for ambulance services}

\author{Pedro A.\ Pury \\
Facultad de Matem\'atica, Astronom\'\i a,
F\'\i sica y Computaci\'on \\
Universidad Nacional de C\'ordoba \\
Ciudad Universitaria, X5000HUA C\'ordoba, Argentina \\
\texttt{pury@famaf.unc.edu.ar}}

%\date{February 21, 2016}
%\date{December 09, 2025)
\date{}
%
%%%%%%%%%%%%%%%%%%%%%%%%%%%%%%%%%%%%%%%%%%%%%%%%%%%%%%%%%%%%%%%%%%%%%%
\maketitle
%%%%%%%%%%%%%%%%%%%%%%%%%%%%%%%%%%%%%%%%%%%%%%%%%%%%%%%%%%%%%%%%%%%%%%
%
\begin{abstract}
We present predictive tools to calculate the number of ambulances
needed according to demand of entrance calls and time of service.
Our analysis discriminates between emergency and non-urgent calls.
First, we consider the nonstationary regime where we apply previous
results of first-passage time of one dimensional random walks.
Then, we reconsider the stationary regime with a detailed discussion
of the conditional probabilities and we discuss the key performance
indicators.
\end{abstract}
%
%%%%%%%%%%%%%%%%%%%%%%%%%%%%%%%%%%%%%%%%%%%%%%%%%%%%%%%%%%%%%%%%%%%%%%

%%%%%%%%%%%%%%%%%%%%%%%%%%%%%%%%%%%%%%%%%%%%%%%%%%%%%%%%%%%%%%%%%%%%%%
\section{Introduction}
\label{sec:intro}
%%%%%%%%%%%%%%%%%%%%%%%%%%%%%%%%%%%%%%%%%%%%%%%%%%%%%%%%%%%%%%%%%%%%%%

Emergency Medical Services (EMS) involve several operational
decisions concerning optimizing the station location deployment
for ambulances and the selection of the number of ambulances
available in the fleet at different moments of the day or
on different days in the week.
Quantitative and predictive tools to assist the ambulance
management are becoming increasingly important in the solution
of economic and medical aspects of the problem~\cite{Gol04}.
Among these tools, modeling health care problems with
queueing theory has gained in significance in the last
sixty years~\cite{LS13}.

Queueing theory has been used in the study of spatial and temporal
distributions of demand of EMS in order of simulate the behavior
of the system.
Particularly, the ambulance location problem has been subject
of considerable attention~(See Refs.~\cite{Hall71,BL74,TWM07}
and references therein) and has been applied to support decision
making~\cite{SD08}.
However, the problem of the number of ambulances required for
service has received little mathematical attention beyond the
prediction by the Little's Law~\cite{Little61, Little11}.
The objective of this work is to derive useful quantities
to predict the number of ambulances needed in real-time operation
of EMS. This is a analytical work but with a didactic approach.
We address here both modeling with clear mathematical derivations
and their significance to operations research.

Our model is basically a call center~\cite{KM02} and can be
described by exponential interarrival and service times,
and $s=M$ servers: $M/M/s/GD/\infty/\infty$~\cite{Win03}.
Thus, our queueing model is based on two average times:
$T_C$, the mean difference between call arrival times and
$T_S$, the mean service time of an individual ambulance.
The service time involves the complete time lapse between the
dispatch of the ambulance from the base and the its release
for subsequent utilization. Thus, the service time sums up
all the transport times of the ambulance plus the specific
time for the medical attention at the scene.
We are interested in the behavior of the system over a finite
time interval, but long compared with $T_C$ and $T_S$~\cite{CS61}.
We focus on the nonstationary regime, where we apply the concept
of the mean first-passage time (MFPT)~\cite{Red01},
as well as in the stationary regime, where we are interested
in the key performance quantities.

The paper is organized in three main sections.
In Section~\ref{sec:RW} we describe the random walk model
and the problem assumptions are emphasized.
Section~\ref{Sec:Critical} focus on the calculation of the time
to the next critical condition, that is when all ambulances are busy,
whereas in Section~\ref{Sec:NonCritical} we discuss standard
results of queueing theory with emphasis in the relevant performance
indicators for the queue of clients and level of service of a home
medical care service. Here, we also provide an alternative proof
of the Little's Law.
The complementary mathematical details of Sections~\ref{Sec:Critical}
and~\ref{Sec:NonCritical} are relegated to corresponding subsections
in the appendix to enhance readability.
Finally, in Section~\ref{sec:fin} we briefly summarize
our main results.

%%%%%%%%%%%%%%%%%%%%%%%%%%%%%%%%%%%%%%%%%%%%%%%%%%%%%%%%%%%%%%%%%%%%%%
\section{The random walk model}
\label{sec:RW}
%%%%%%%%%%%%%%%%%%%%%%%%%%%%%%%%%%%%%%%%%%%%%%%%%%%%%%%%%%%%%%%%%%%%%%

Formally, we represent the system by the total number $M$
of ambulances in the service and by the number $n$ of received
calls not served yet. $n$ is the state of occupation of the system.
When $n=0$, all the ambulances are in the base and there are
not calls in queue.
For $0 < n < M$, there are not waiting calls and $n$ ambulances
are in course of action. That is, in transit from base to the call
location, attending at patient location, or in transit to the hospital.
In an equivalent way, we can say that there are $n$ patients
simultaneously being served.
When $n=M$, the system is at the critical state. Even though there
are not calls in waiting, all the servers have been assigned
to calls and, in consequence, there are not any ambulance available
to serve an eventual next incoming call.
For $n>M$, the system is saturated. All ambulances are occupied
and there are $(n-M)$ calls waiting to be served in the queue.

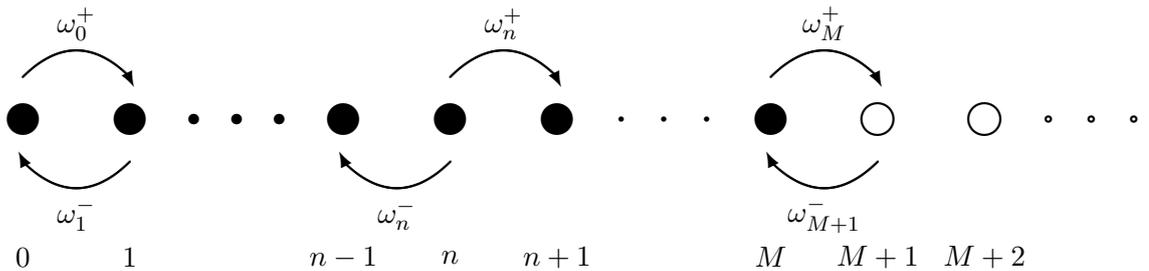
\begin{figure}[!hpt]
\begin{center}
\unitlength=0.80pt
\thicklines
\begin{picture}(500.00,100.00)(0.00,0.00)
\put(05.00,50.00){\circle*{15.00}}
\put(55.00,50.00){\circle*{15.00}}
\put(85.00,50.00){\circle*{5.00}}
\put(105.00,50.00){\circle*{5.00}}
\put(125.00,50.00){\circle*{5.00}}
\put(155.00,50.00){\circle*{15.00}}
\put(205.00,50.00){\circle*{15.00}}
\put(255.00,50.00){\circle*{15.00}}
\put(285.00,50.00){\circle*{3.00}}
\put(305.00,50.00){\circle*{3.00}}
\put(325.00,50.00){\circle*{3.00}}
\put(355.00,50.00){\circle*{15.00}}
\put(405.00,50.00){\circle{15.00}}
\put(455.00,50.00){\circle{15.00}}
\put(485.00,50.00){\circle{3.00}}
\put(505.00,50.00){\circle{3.00}}
\put(525.00,50.00){\circle{3.00}}
\put(5.00,-15.00){\makebox(0.00,0.00){$0$}}
\put(55.00,-15.00){\makebox(0.00,0.00){$1$}}
\put(155.00,-15.00){\makebox(0.00,0.00){$n-1$}}
\put(205.00,-15.00){\makebox(0.00,0.00){$n$}}
\put(255.00,-15.00){\makebox(0.00,0.00){$n+1$}}
\put(355.00,-15.00){\makebox(0.00,0.00){$M$}}
\put(405.00,-15.00){\makebox(0.00,0.00){$M+1$}}
\put(455.00,-15.00){\makebox(0.00,0.00){$M+2$}}
\qbezier(5.00,70.00)(30.00,95.00)(55.00,70.00)
\put(57.00,65.00){\vector(1,-2){0.00}}
\qbezier(205.00,70.00)(230.00,95.00)(255.00,70.00)
\put(257.00,65.00){\vector(1,-2){0.00}}
\qbezier(355.00,70.00)(380.00,95.00)(405.00,70.00)
\put(407.00,65.00){\vector(1,-2){0.00}}
\qbezier(5.00,30.00)(30.00,5.00)(55.00,30.00)
\put(3.00,35.00){\vector(-1,2){0.00}}
\qbezier(155.00,30.00)(180.00,5.00)(205.00,30.00)
\put(153.00,35.00){\vector(-1,2){0.00}}
\qbezier(355.00,30.00)(380.00,5.00)(405.00,30.00)
\put(353.00,35.00){\vector(-1,2){0.00}}
\put(30.00,95.00){\makebox(0.00,0.00){$\omega^+_0$}}
\put(230.00,95.00){\makebox(0.00,0.00){$\omega^+_n$}}
\put(380.00,95.00){\makebox(0.00,0.00){$\omega^+_M$}}
\put(30.00,5.00){\makebox(0.00,0.00){$\omega^-_1$}}
\put(180.00,5.00){\makebox(0.00,0.00){$\omega^-_n$}}
\put(380.00,5.00){\makebox(0.00,0.00){$\omega^-_{M+1}$}}
\end{picture}

\vspace{25pt}
\caption{Scheme of transition probabilities between system's
states in the upper side and the state of occupation of the system
in the bottom line, for a system with $M$ servers.}
\label{RWmodel}
\end{center}
\end{figure}

At any time, the system can change its state between its nearest
neighbors. Thus, the transition probability per unit time from
the state $n$ to $(n+1)$  is denoted by $\omega^+_n$, whereas
the transition probability toward the lower occupation state
is given by $\omega^-_n$.
In Figure~\ref{RWmodel}, we sketch the possible transitions
for the system.

%%%%%%%%%%%%%%%%%%%%%%%%%%%%%%%%%%%%%%%%%%%%%%%%%%%%%%%%%%%%%%%%%%%%%%
%%%%%%%%%%%%%%%%%%%%%%%%%%%%%%%%%%%%%%%%%%%%%%%%%%%%%%%%%%%%%%%%%%%%%%

The dynamics of the probability $P_n(t)$ of finding the system
at time $t$ in the state $n$ is ruled by the master equations
of a random walk between nearest-neighbor sites with a reflecting
boundary at the origin, that is a birth--death process~\cite{Win03},
\begin{equation}
\begin{array}{lcl}
\displaystyle\frac{dP_0(t)}{dt} &=&
\omega^-_1 \,P_1(t) -\omega^+_0 \,P_0(t) \,,
\\
\rule{0cm}{0.7cm}
\displaystyle\frac{dP_n(t)}{dt} &=& \omega^-_{n+1} \,P_{n+1}(t)
+\omega^+_{n-1} \,P_{n-1}(t) - (\omega^+_n + \omega^-_n) \,P_n(t)
\;\; \mbox{$(n \geq 1)$} \,.
\end{array}
\label{mastereq}
\end{equation}
We restrict our analysis to the case in which the times between
calls and the service times are {\em independent exponential}
random variables.
The assumption that call arrival volume per unit of time is Poisson
distributed is an standard choice in industry~\cite{MMWH11}.
Thus, the time between calls is an exponential variable with
rate or mean number of calls per unit time $\lambda = 1/T_C$.
On the other hand, the service time rate or mean number
of attention per unit time and per ambulance results $\mu = 1 / T_S$.
Therefore, the transition probabilities are defined by~\cite{CS61}
% D.R. Cox and W.L. Smith, Queues, Sec.2.4 (iii)
%
\begin{equation}
\begin{array}{l}
\omega^+_n = \lambda \,, \;\forall n \,, \\
\omega^-_n =
\left\{
\begin{array}{ll}
n \, \mu  & (n \leq M) \,, \\
M \, \mu  & (n \geq M) \,.
\end{array}
\right.
\rule{0cm}{0.7cm}
\end{array}
\label{Mservers}
\end{equation}
In this manner, only $\omega^-_n$ depends on the state of the system.
The average times $T_C$ and $T_S$ are all the experimental information
needed to characterize the problem.

%%%%%%%%%%%%%%%%%%%%%%%%%%%%%%%%%%%%%%%%%%%%%%%%%%%%%%%%%%%%%%%%%%%%%%
\section{The critical emergency problem}
\label{Sec:Critical}
%%%%%%%%%%%%%%%%%%%%%%%%%%%%%%%%%%%%%%%%%%%%%%%%%%%%%%%%%%%%%%%%%%%%%%

The ambulance industry has the goal of provide care within 8 minutes
for heart attack and cardiac arrest~\cite{EBH79} and major
trauma~\cite{FHSI95} (for critical considerations see Ref.~\cite{PM02}).
Usually, the time lapse between picking the call up and the
arrival of the ambulance at the scene is mainly consumed in transport.
Thus, to achieve this goal is necessary to respond the calls
immediately, without putting any EMS call in queue.
Therefore, the most important aspect of emergency medical management
is avoiding the saturation of the system. The prediction
of the critical condition ($n=M$) is the particular interest
for the quality of the medical service as much as the economic
management of the service, given that the critical condition strongly
depends on the number $M$ of ambulances simultaneously in service.

The mean time to the next critical condition is function of the
initial state of the system, and can be calculated as the MFPT
of a one dimensional random walk to site $M$ with a reflecting
boundary at the opposite extreme (see Figure~\ref{RWmodel}).
For the critical problem, the initial condition $n$
is restricted to the values in the interval
$[ 0, \dots, M]$. MFPT is a dynamical variable
that can not be computed in the steady-state.
Using known results in the literature~\cite{PC03},
the mean time to the critical condition as function
of the initial state of the system results
\begin{equation}
\begin{array}{lcl}
T(0) &=&
T_C \left(
M+1 + \displaystyle\sum_{k=0}^{M-1} \,
\frac{\gamma^{-k}}{k!} \sum_{i=k+1}^{M} i! \,\gamma^i
\right) \;,
\\
T(1) &=& T(0) - T_C \;,
\\
T(n) &=& T(0) - T_C \left(
n + \displaystyle\sum_{k=0}^{n-2} \,
\frac{\gamma^{-k}}{k!} \sum_{i=k+1}^{n-1} i! \,\gamma^i
\right)
\;\; (2 \leq n \leq M)\,,
\label{MFPT:critical}
\end{array}
\end{equation}
where the parameter $\gamma = \mu / \lambda = T_C/T_S$.
The mathematical details are given in Appendix~\ref{sec:MFPT}.

\begin{figure}[ht]
\begin{center}
\includegraphics[clip,width=0.85\textwidth]{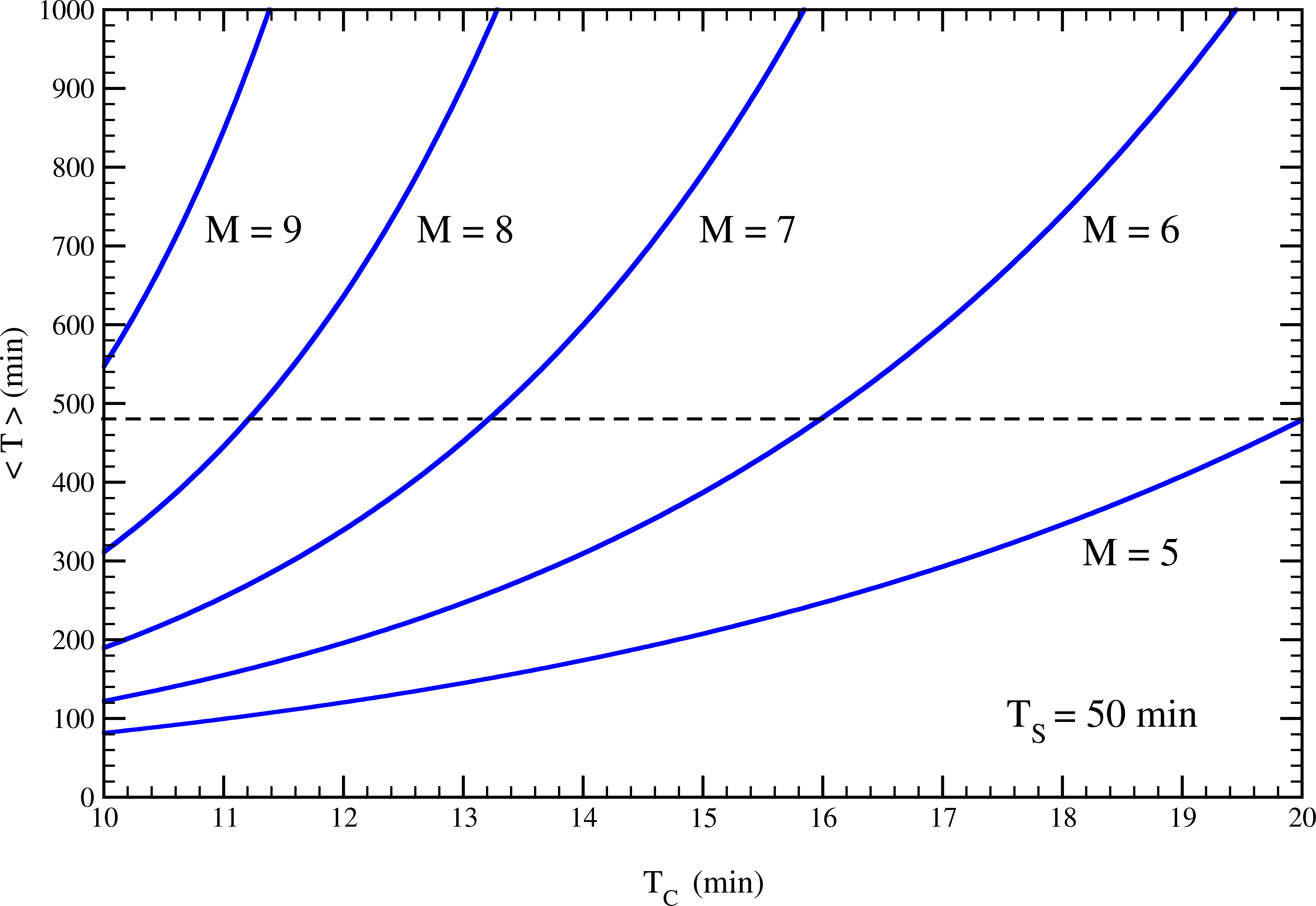}
\caption{Mean time to critical condition as function of
the mean time between incoming calls.}
\label{f-mfpt}
\end{center}
\end{figure}

Particularly, $T(M)$ is the mean time between two critical
conditions of the system, but without reaching saturation.
Furthermore, as we want a quantity independent of the initial state,
we average over $n = 0, \ldots, M$. For this purpose, we define
\begin{equation}
<T> = \displaystyle\frac{1}{M+1} \,\sum_{n=0}^M \,T(n) \,.
\label{<T>}
\end{equation}
In this way, knowing the values of $T_C$ and $T_S$,
the expressions of Eq.~(\ref{MFPT:critical}) can be
numerically evaluated in a very direct way.
In Figure~\ref{f-mfpt} we show the plots of $<T>$, according
to Eqs.~(\ref{MFPT:critical}) and~(\ref{<T>}),
as function of the mean time between calls $T_C$.
We have sketched a characteristic situation where the mean service
time is $50\,$min and we considered the number of ambulances
$M=5, \dots, 9$. The curves clearly show the non-linear behavior
of $<T>$.

Emergency medical care is an activity brought completely
under protocol and in consequence the time of paramedic
duties on scene have little variance.
However, in a modern metropolis, transport time could
consume an important part of the total service time
and it have an specific hourly pattern.
But, in practical situations, $T_S$ has small variations
over the day in comparison with $T_C$. The fluctuations
of the rate of entrance calls, $\lambda$, have a defined
pattern with marked differences during the day.
Usually, the mean number of EMS calls per hour remains high
during a time span of approximately eight hours between noon
and 8:00 p.m.~\cite{MMWH11}. This distinct shape is similar
for all days of the week.

Considering that $T_C$ is measured in the time interval of highest
number of calls, the mean time to critical condition must be longer
that this time span to ensure the service's quality and avoiding
saturation.
For example, in Figure~\ref{f-mfpt} we see that the system do not
reach the critical condition within the interval of eight hours
of high $\lambda$ (horizontal dashed line) for $T_C>16\,$min
with six ambulances, and for $T_C>13.2\,$min with seven ambulances.

In a similar way, to optimize the number of ambulances simultaneously
in service, our analysis can be reproduced for any time interval
of the day, for different days of the week, and at different
locations of ambulance bases.
Usually, the call rate $\lambda$, measured in a given interval
during the day, is a linear function of the numbers of
affiliates of the EMS but with seasonal fluctuations.
This estimation gives us a reference for the average
value of $T_C$ in the long term.
In the short run, using the forecasting of the volume of call
arrival~\cite{SSP09,MMWH11}, our analysis allows us to estimate
the diary and hourly demand of ambulances.
Thus, we can easily design a simple and precise predictive tool
for the fleet of ambulances required for a EMS by combining
the use of estimation or forecasting of $T_C$ with our analysis
of Figure~\ref{f-mfpt}, tailored for each particular case.

%%%%%%%%%%%%%%%%%%%%%%%%%%%%%%%%%%%%%%%%%%%%%%%%%%%%%%%%%%%%%%%%%%%%%%
\section{The non-urgent call problem}
\label{Sec:NonCritical}
%%%%%%%%%%%%%%%%%%%%%%%%%%%%%%%%%%%%%%%%%%%%%%%%%%%%%%%%%%%%%%%%%%%%%%

Most of the ambulance service providers, besides dealing with emergency
calls, also bring medical home assistance for non-urgent calls.
Usually, if a call is evaluated as non-urgent, the dispatcher
derive it to another system to alleviate the use of emergency
ambulances. In this case, contrary to the emergency management,
non-urgent calls are allowed occasionally to be driven under
saturation ($n>M$).
Thus, the length of the queue and the waiting time of the patients
are the quantities of interest for the ongoing process. However,
the system is driven in steady-state only under a special condition.

%%%%%%%%%%%%%%%%%%%%%%%%%%%%%%%%%%%%%%%%%%%%%%%%%%%%%%%%%%%%%%%%%%%%%
\subsection{Steady-state}
%%%%%%%%%%%%%%%%%%%%%%%%%%%%%%%%%%%%%%%%%%%%%%%%%%%%%%%%%%%%%%%%%%%%%

\begin{figure}[ht]
\begin{center}
\includegraphics[clip,width=0.75\textwidth]{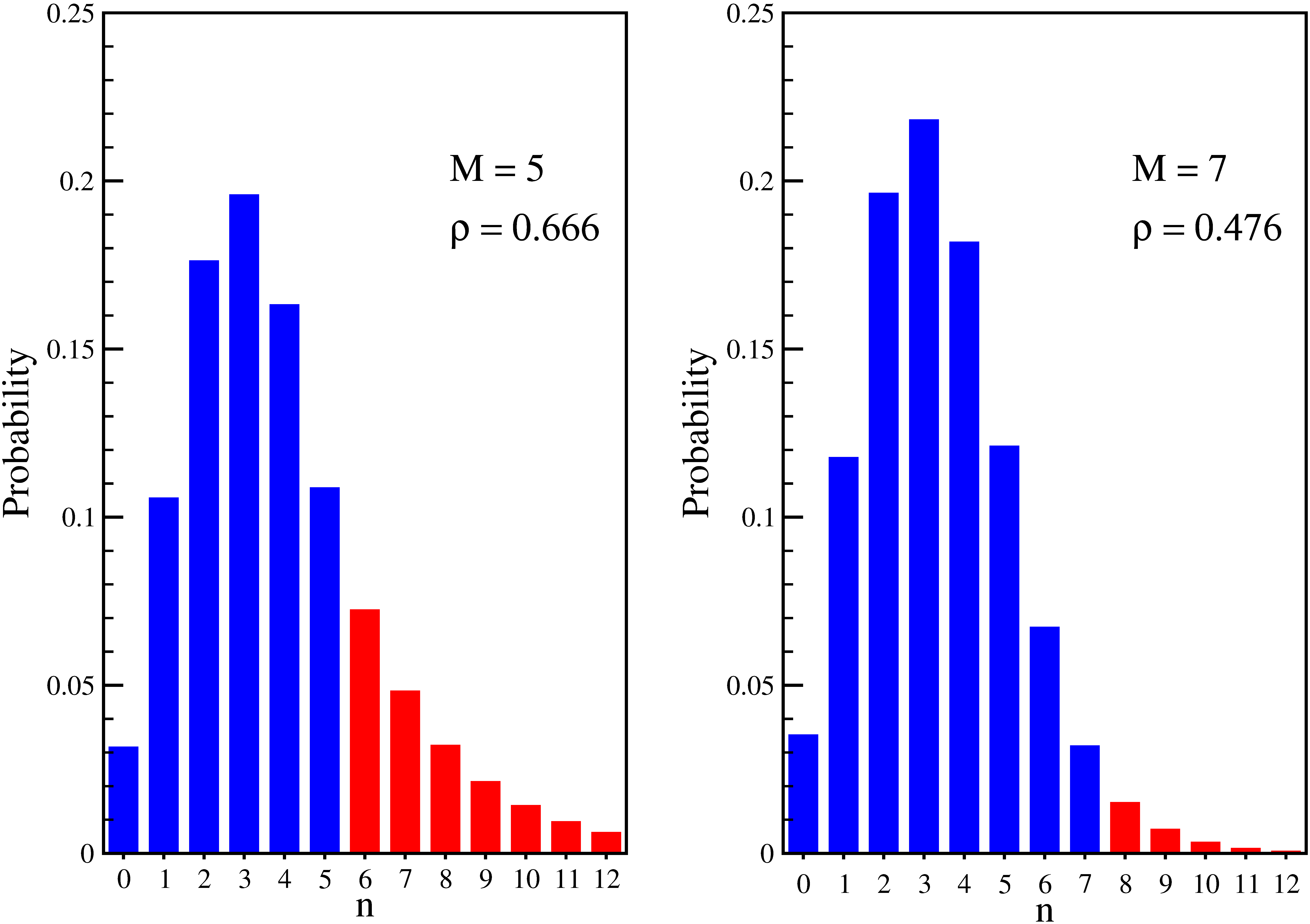}
\caption{Probability distribution in the steady-state for
$T_S = 50\,$min and $T_C = 15\,$min.}
\label{f-probs}
\end{center}
\end{figure}

First, we define the dimensionless control parameter
\begin{equation}
\rho = \displaystyle \frac{\lambda}{M \,\mu} \,,
\label{rho}
\end{equation}
which is also called {\em traffic intensity} in queueing
theory~\cite{CS61}.
The system has stationary state if and only if $\rho  < 1$.
Under this condition, the limit
$\pi_n = \lim_{t \rightarrow \infty} P_n(t)$
exists and according to the analysis of Appendix~\ref{sec:SS}
is given by
\begin{equation}
\pi_n = \displaystyle \frac{1}{S} \,\left\{
\begin{array}{ll}
\displaystyle \frac{M^n}{n!} \,\rho^n
& (0 \leq n \leq M) \,, \\
\rule{0cm}{0.8cm}
\displaystyle \frac{M^M}{M!} \,\rho^n
& (n \geq M) \,,
\end{array}
\right.
\label{PnM}
\end{equation}
where
\begin{equation}
S = \displaystyle\sum_{n=0}^{M-1} \frac{(M \,\rho)^n}{n!}
+ \displaystyle\frac{(M \,\rho)^M}{M! \,(1-\rho)} \,.
\label{S}
\end{equation}
In Figure~\ref{f-probs} we show plots for the probability
distribution, according to Eqs.~(\ref{PnM}) and~(\ref{S}).
Left panel corresponds to a fleet of five ambulances,
whereas the right panel is for seven servers.

When the condition $\rho  < 1$ is not fulfilled, the probabilities
$\pi_n$ are not defined. This situation corresponds to the collapse
of the system when it is unable to deal with calls as fast as they
arrive and the queue of waiting calls will develop without limit.

%%%%%%%%%%%%%%%%%%%%%%%%%%%%%%%%%%%%%%%%%%%%%%%%%%%%%%%%%%%%%%%%%%%%%%
\subsection{Queue length}
%%%%%%%%%%%%%%%%%%%%%%%%%%%%%%%%%%%%%%%%%%%%%%%%%%%%%%%%%%%%%%%%%%%%%%

If a call arrives when all servers are busy, it will be put in queue.
The probability of full occupation of servers results
\begin{equation}
P(\mbox{occup}) = P(n \geq M) = \sum_{n=M}^{\infty} \pi_n =
\frac{1}{S} \,\frac{M^M}{M!} \sum_{n=M}^{\infty} \rho^n \,,
\label{Pn>M}
\end{equation}
and under the condition $\rho<1$, using Eq.~(\ref{seriesg}),
we obtain
\begin{equation}
P(\mbox{occup}) = \displaystyle
\frac{(M \,\rho)^M}{M! \;(1-\rho) \,S} \,.
\label{occup}
\end{equation}
The red tails in Figure~\ref{f-probs} are the probabilities
of full occupation for $M=5$ (left) and $M=7$ (right)
ambulances, respectively.

To calculate the length of the genuine queue formed when the $M$
servers are busy, we need the conditional probability that the
system is at state $n$, given that all servers are occupied.
From Eq.~(\ref{PnM}) (for $n \geq M$) and~(\ref{occup}) we obtain
\begin{equation}
P(n|\mbox{occup}) = \rho^{n-M} \,(1-\rho) \qquad (n \geq M) \,.
\label{condcional}
\end{equation}
Setting $n=k+M$, $k=0,1,\dots$ results that the conditional
probability of having $k$ calls in the waiting row under
full occupation is
\begin{equation}
P(k| \mbox{occup}) = \rho^k \,(1-\rho) \qquad (k \geq 0) \,.
\label{geom}
\end{equation}
This is the geometric distribution with parameter $1-\rho$.
Therefore, the average length of the genuine queue,
$\left< L \right> = \sum_{k=0}^{\infty} \,k \,P(k| \mbox{occup})$,
and its standard deviation, $\sigma_L$, result
\begin{equation}
\left< L \right> = \frac{\rho}{1-\rho} \,,\qquad
\sigma_L = \frac{\sqrt{\rho}}{(1-\rho)} \,.
\label{cola}
\end{equation}
Both quantities evidently diverge in the limit $\rho \rightarrow 1$.
\begin{figure}[ht]
\begin{center}
\includegraphics[clip,width=0.80\textwidth]{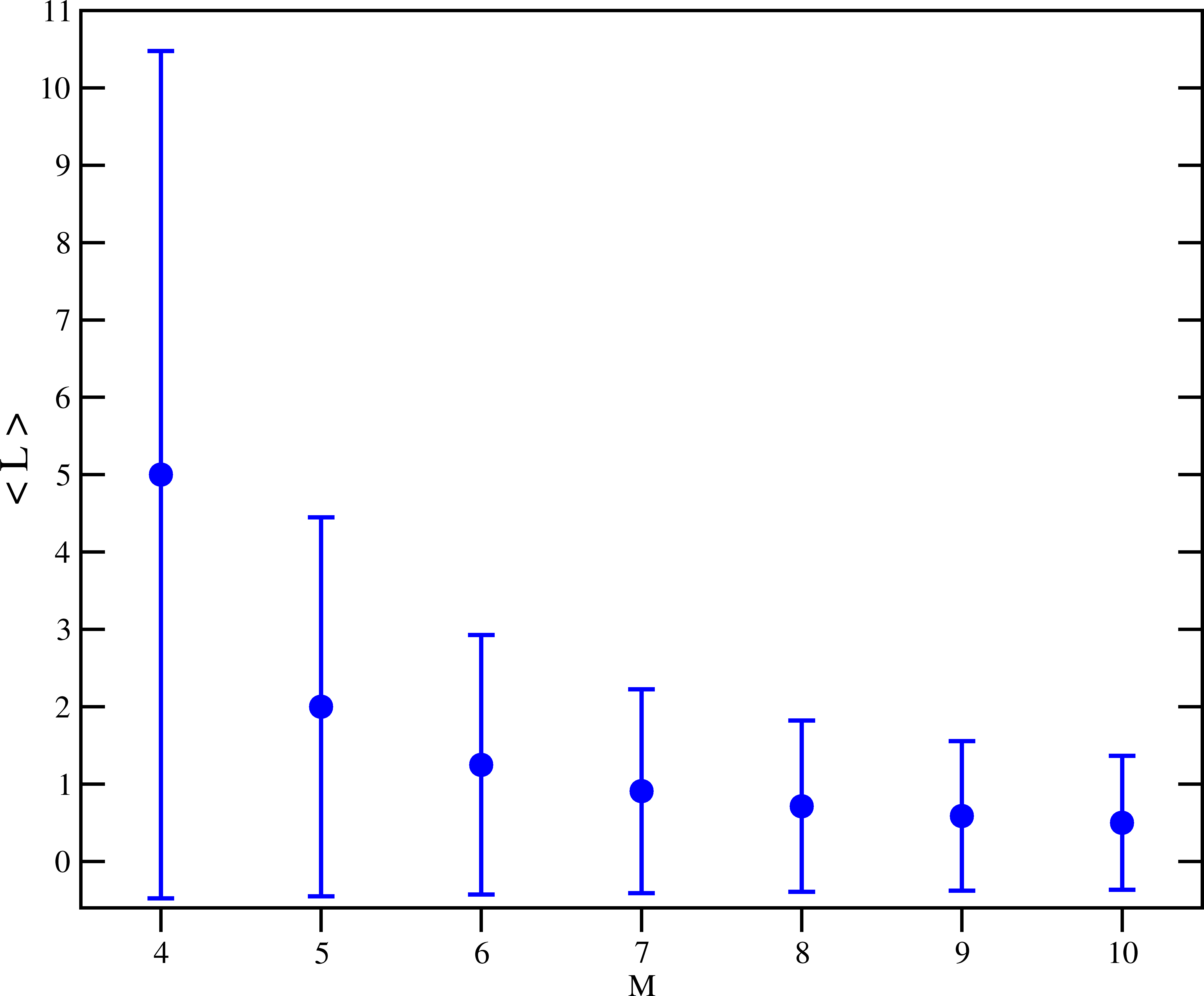}
\caption{Queue length for $T_S = 50\,$min and $T_C = 15\,$min.}
\label{f-length}
\end{center}
\end{figure}
To visualize the effect of the number of ambulances,
we plot in Figure~\ref{f-length} the expressions of Eq.~(\ref{cola})
for $M = 4 \,(\rho=0.833), \dots, 10 \,(\rho=0.333)$
In this example, for $M < 4$ results $\rho > 1$.
We can see that to increase the fleet in more than
six ambulance has not any practical consequence
in the queue length.

%%%%%%%%%%%%%%%%%%%%%%%%%%%%%%%%%%%%%%%%%%%%%%%%%%%%%%%%%%%%%%%%%%%%%%%
\subsection{Quality of service}
%%%%%%%%%%%%%%%%%%%%%%%%%%%%%%%%%%%%%%%%%%%%%%%%%%%%%%%%%%%%%%%%%%%%%%%

Now, we consider the arrival of a new call when the state
of the system is $n=M+k$, i.e., all servers are busy
and there are $k \geq 0$ calls in waiting.
Under the queue discipline {\em first-come, first-served},
the time that the new patient will have to wait until an
ambulance will be dispatched to his or her location, will
be the sum of the waiting times of $N=k+1$ patients:
The first $k$ in the row plus the time of any of the
patients in service at the arrival time.

The probability distribution of the sum of $N$ exponential
independent random variables with parameter $\alpha$
is a Gamma distribution with parameters $(N,\alpha)$~\cite{Ross06},
\begin{equation}
f(t;N,\alpha) = \alpha \,e^{-\alpha \,t}
\,\frac{(\alpha \,t)^{N-1}}{(N-1)!} \,.
\label{Gamma}
\end{equation}
Thus, the conditional density of probability that a patient
will be waiting a time $t$ given $k$ patients ahead in the row
is given by Eq.~(\ref{Gamma}) with $\alpha = M \,\mu$ and $N=k+1$.
We also know that the probability of having $k$ patients in the
row is given by Eq.~(\ref{geom}). Therefore, the probability
density function of the waiting time in the row results
\begin{equation}
g(t) = \sum_{k=0}^{\infty} (1-\rho) \,\rho^k \,f(t;(k+1),M \,\mu) \,.
\label{tespera}
\end{equation}
Summing up the series,
\begin{equation}
\sum_{k=0}^{\infty} \frac{(\alpha \,\rho \,t)^k}{k!}
= e^{\alpha \,\rho \,t},
\end{equation}
allows recast Eq.~(\ref{tespera}) as
\begin{equation}
g(t) = (1-\rho) \,\alpha \,e^{-(1-\rho) \,\alpha \,t} \,.
\label{exponencial1}
\end{equation}
Thus, we obtain an exponential distribution with parameter
$(1-\rho) \,\alpha = (1-\rho) \,M \,\mu$.
In this manner, when all servers are busy, the mean
waiting time of a patient in the row before being served is
\begin{equation}
\left< T \right> = \displaystyle\frac{1}{(1-\rho) \,M \,\mu}
= \displaystyle\frac{\left< L \right>}{\lambda} \,.
\label{tmedio}
\end{equation}
The result $\left< L \right> = \lambda \,\left< T \right>$
is the expression of the well known Little's Law
\cite{Little61, Little11}. Our derivation is an alternative
statistical approach in the stationary framework~\cite{KW13}.

\begin{figure}[ht]
\begin{center}
\includegraphics[clip,width=0.80\textwidth]{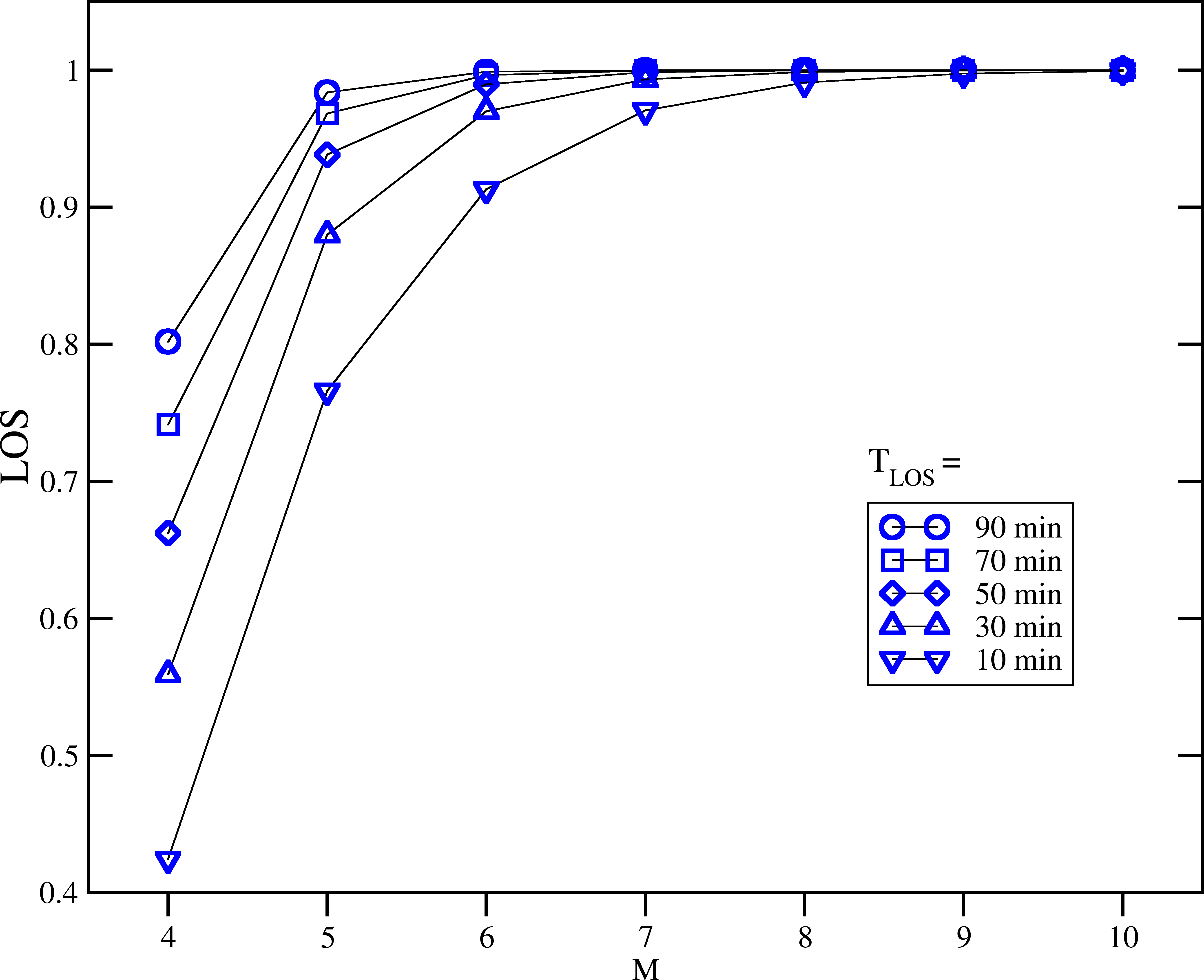}
\caption{Level of service for $T_S = 50\,$min and $T_C = 15\,$min
and different values of $T_{\mbox{\tiny LOS}}$.
Lines between points are only to guide the eye.}
\label{f-LOS}
\end{center}
\end{figure}

We define the level of service (LOS) as the fraction of
patients served in a time less than a predefined threshold
of quality $T_{\mbox{\tiny LOS}}$.
When a new call arrives, if there are idle servers, an ambulance
is dispatched and there is not waiting time, but under full
occupation, the quality threshold is fulfilled with probability
\begin{equation}
P(t<T_{\mbox{\tiny LOS}}) = \int_0^{T_{\mbox{\tiny LOS}}} \,g(t) dt
= 1 - \,e^{-(1-\rho) \,\alpha \,T_{\mbox{\tiny LOS}}} \,.
\label{t<T}
\end{equation}
Thus, $\mbox{LOS} = (1-P(\mbox{occup})) \,1
+ P(\mbox{occup}) \,P(t<T_{\mbox{\tiny LOS}})
=  1 - P(\mbox{occup}) \,(1-P(t<T_{\mbox{\tiny LOS}}))$,
and in our case we obtain,
\begin{equation}
\mbox{LOS} =
1 - P(\mbox{occup}) \,e^{-(1-\rho) \,M\,\mu \,T_{\mbox{\tiny LOS}}} \,,
\label{LOS}
\end{equation}
where $P(\mbox{occup})$ is given by Eq.~(\ref{occup}).
Then, $\mbox{LOS} =  1$ if and only if $P(\mbox{occup}) = 0$
which is the desired condition for management of emergency calls.
In Figure~\ref{f-LOS}, we illustrate the dependence of LOS
in the number of ambulances.
%

%%%%%%%%%%%%%%%%%%%%%%%%%%%%%%%%%%%%%%%%%%%%%%%%%%%%%%%%%%%%%%%%%%%%%%
\subsection{Performance of servers}
%%%%%%%%%%%%%%%%%%%%%%%%%%%%%%%%%%%%%%%%%%%%%%%%%%%%%%%%%%%%%%%%%%%%%%

\begin{figure}[ht]
\begin{center}
\includegraphics[clip,width=0.80\textwidth]{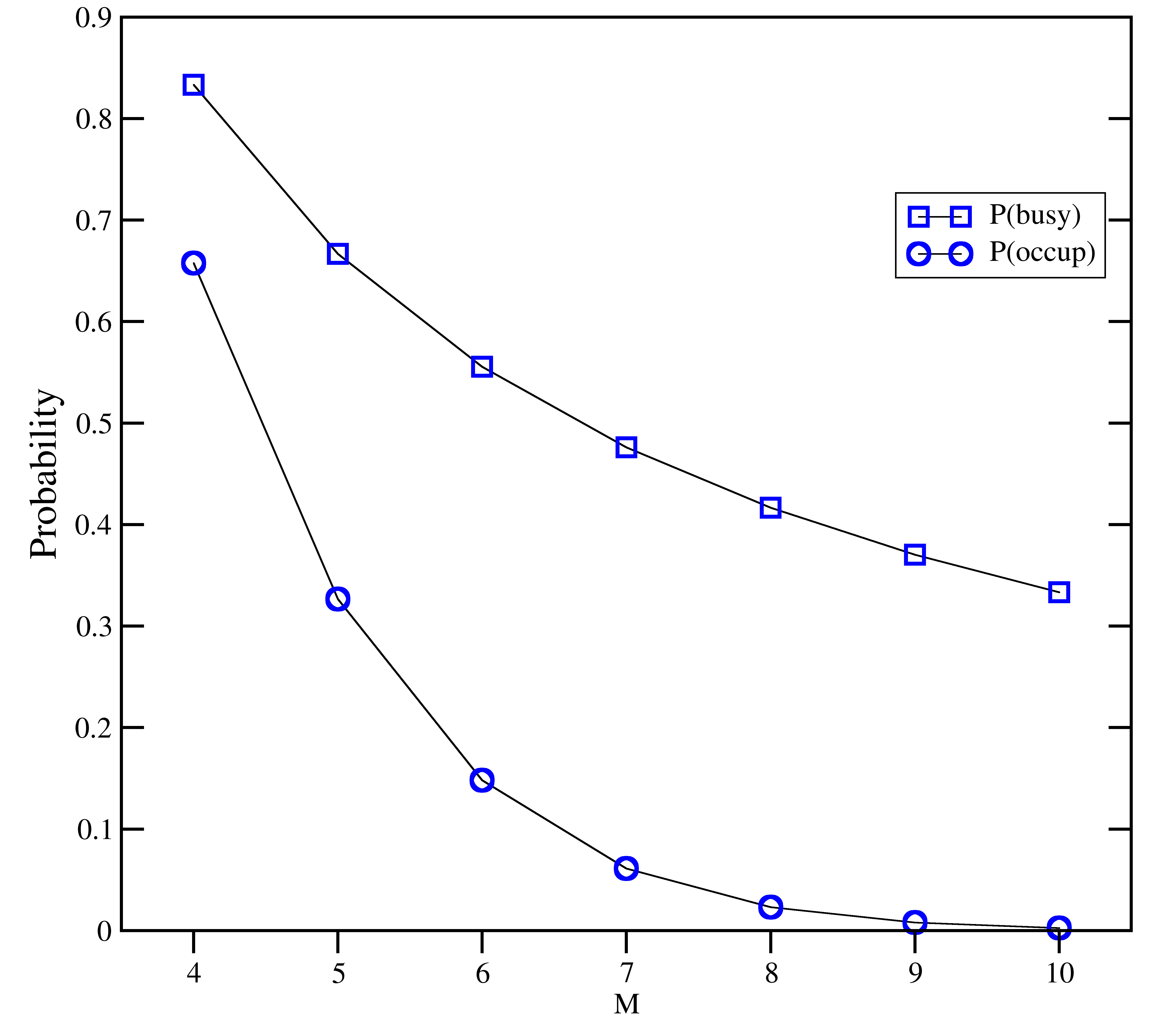}
\caption{$P(\mbox{busy})$ and $ P(\mbox{occup})$
for $T_S = 50\,$min and $T_C = 15\,$min.
Lines between points are only to guide the eye.}
\label{f-poccsb}
\end{center}
\end{figure}

\begin{figure}[ht]
\begin{center}
\includegraphics[clip,width=0.80\textwidth]{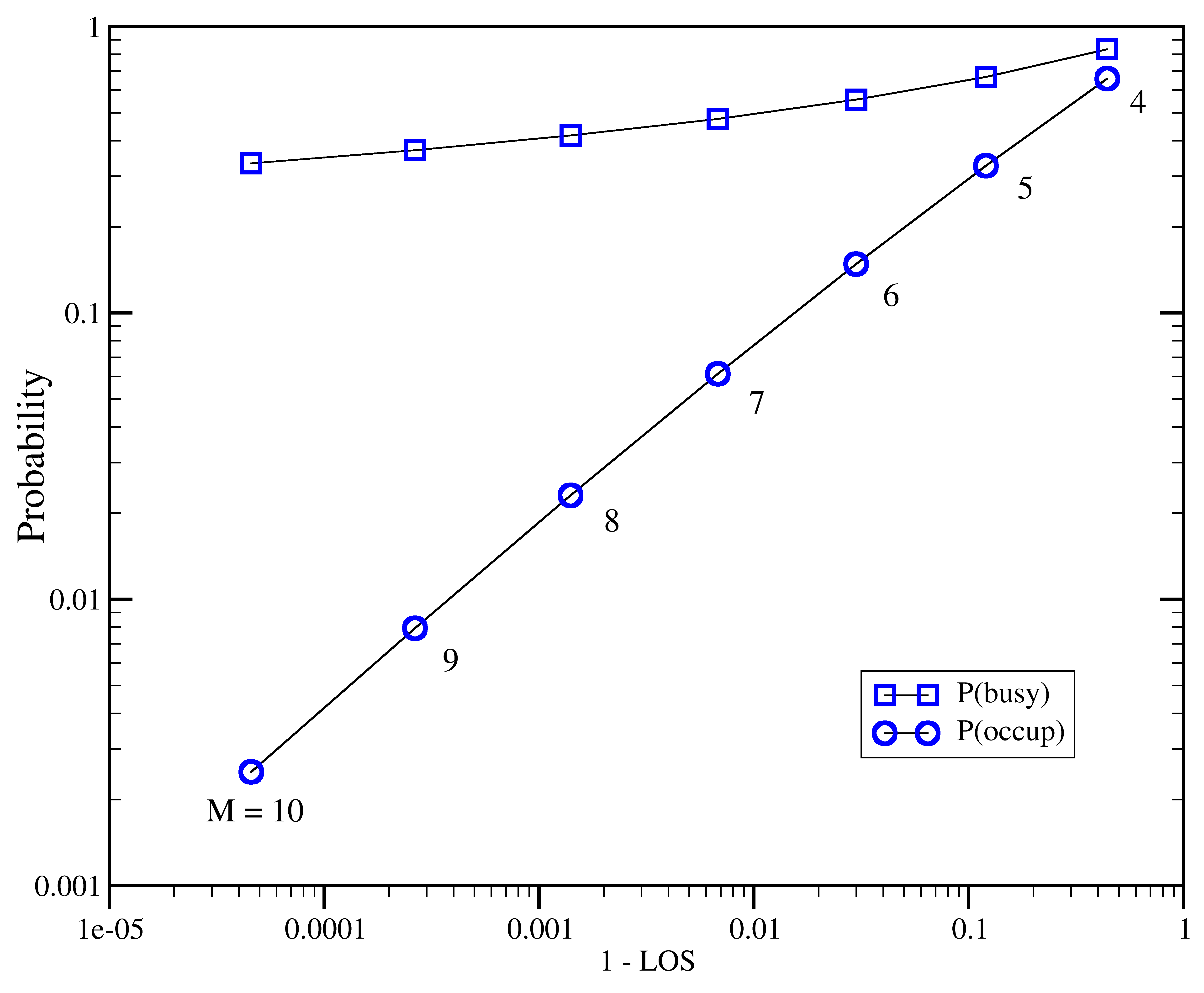}
\caption{Log-log plane of probabitity ($P(\mbox{busy})$
or $ P(\mbox{occup})$) and $(1 - \mbox{LOS})$ as function
of $M$ for $T_S = 50\,$min, $T_C = 15\,$min
and,  $T_{\mbox{\tiny LOS}} = 30\,$min.
Lines between points are only to guide the eye.}
\label{f-tradeoff}
\end{center}
\end{figure}

The probability that a given server is busy can be written as
\begin{equation}
P(\mbox{busy}) = \sum_{n=1}^{\infty} \,P(\mbox{busy}|n) \,\pi_n \,,
\label{so_def}
\end{equation}
where $\pi_n$ is given by Eq.~(\ref{PnM}) and $P(\mbox{busy}|n)$
is the conditional probability of a server busy given that
the state of the system is $n$.
For $n<M$ and under the assumption that the assignment of calls
to servers is at random if there is more than one idle,
using simple combinatorial calculation, results
\begin{equation}
P(\mbox{busy}|n) =
\frac{\displaystyle {{M-1}\choose{n-1}}}
     {\displaystyle {{M}\choose{n}}}
= \frac{n}{M}\,,
\label{so|n}
\end{equation}
whereas, for $n \geq M$, $P(\mbox{busy}|n) = 1$. Then,
\begin{equation}
P(\mbox{busy}) = \sum_{n=1}^{M-1} \frac{n}{M} \,\pi_n + P(\mbox{occup})\,.
\label{busy}
\end{equation}
Therefore, from Eqs.~(\ref{PnM}) and~(\ref{occup}) results
\begin{equation}
P(\mbox{busy}) = \displaystyle\frac{1}{S} \left(
\sum_{n=1}^{M-1} \displaystyle\frac{M^{n-1}}{(n-1)!} \,\rho^n
+ \displaystyle\frac{M^M}{M!} \,\frac{\rho^M}{(1-\rho)} \right) \,.
\label{Psb}
\end{equation}
Expression~(\ref{Psb}) can be further simplified, yielding
$P(\mbox{busy}) = \rho$, as expected in the steady-state.
Given the stationary state, $P(\mbox{busy})$ represents
the fraction of time that a given server remains busy.
The dependence on $M$ of $P(\mbox{busy})$ and $ P(\mbox{occup})$
is shown in Figure~\ref{f-poccsb} for our example.
In this situation, from Figure~\ref{f-LOS} and~\ref{f-poccsb},
we can see that a fleet of six ambulances implies LOS greater
than $90 \%$ even though we choose small values of
$T_{\mbox{\tiny LOS}}$.
However, a fleet of six ambulances remains completely occupied
only $14.8 \%$ of time and each server is busy only $55.6 \%$
of time.
To put in clear the trade-off between level of service and use
of resources, we plot in Figure~\ref{f-tradeoff} the log-log plane
of probability ($P(\mbox{busy})$ or $P(\mbox{occup})$) and
$(1 - \mbox{LOS})$ as function of $M$. The graph is constructed
for the fixed value $T_{\mbox{\tiny LOS}} = 30\,$min.

Alternatively, we can look at the mean number of medical attentions
given by the system per unit of time,
\begin{equation}
\left< a \right> = \sum_{n=1}^{\infty} \,\omega^-_n \,\pi_n \,.
\label{atenciones}
\end{equation}
From Eqs.~(\ref{Mservers}) and~(\ref{Pn>M}), we obtain
\begin{equation}
\left< a \right> = \mu \left(
\sum_{n=1}^{M-1} n \,\pi_n + M \,P(\mbox{occup}) \right)
= \mu \,M \,P(\mbox{busy}) \,.
\label{a=betaMPso}
\end{equation}
In this way, if all servers are equivalents,
the simple unit hour utilization,
%http://theemsleader.com/understanding-unit-hour-utilization/
is given by $\mu \,\rho$.
Last, but not least, if server's cost per attention is $C$,
the cost per ambulance in the same unit of time results
$C \,\mu \,\rho$.

%%%%%%%%%%%%%%%%%%%%%%%%%%%%%%%%%%%%%%%%%%%%%%%%%%%%%%%%%%%%%%%%%%%%%%
\section{Concluding remarks}
\label{sec:fin}
%%%%%%%%%%%%%%%%%%%%%%%%%%%%%%%%%%%%%%%%%%%%%%%%%%%%%%%%%%%%%%%%%%%%%%

This work considers the mathematical problem of the number
of ambulances needed in operation for a EMS according to
the mean time between entrance calls and service times.
We developed our analysis in the framework of the queueing theory
for the nonstationary as well as for the stationary regime.

In Section~\ref{Sec:Critical} we presented a novel use of a previous
result of MFPT for calculating the average time to the next critical
condition of full occupation of servers. Our description allows
to cope with the problem of the number of ambulances needed
to avoid that condition.
In Section~\ref{Sec:NonCritical} we rederived standard results
of queueing theory for the stationary state emphasizing the use of
conditional probabilities in the saturation regime when all servers
are busy. We have paid special attention to the key performance
indicators as queue length, level of service, and fraction
of time of use of servers. Our analysis allows stress in simple
mathematical terms the trade-off between quality of service
and use of servers as function of fleet size.

%%%%%%%%%%%%%%%%%%%%%%%%%%%%%%%%%%%%%%%%%%%%%%%%%%%%%%%%%%%%%%%%%%%%%%
\section{Appendix: Mathematics of the random walk}
\label{sec:app}
%%%%%%%%%%%%%%%%%%%%%%%%%%%%%%%%%%%%%%%%%%%%%%%%%%%%%%%%%%%%%%%%%%%%%%

%%%%%%%%%%%%%%%%%%%%%%%%%%%%%%%%%%%%%%%%%%%%%%%%%%%%%%%%%%%%%%%%%%%%%%
\subsection{MFPT}
\label{sec:MFPT}
%%%%%%%%%%%%%%%%%%%%%%%%%%%%%%%%%%%%%%%%%%%%%%%%%%%%%%%%%%%%%%%%%%%%%%

For asymmetric and site dependent transition probabilities,
the analytical expressions for the MFPT of a random walk with
a reflecting boundary condition, as shown in Figure~\ref{RWmodel},
is given by~\cite{PC03},
\begin{equation}
\begin{array}{lcl}
T(0) &=&
\displaystyle\sum_{k=0}^{M} \,\frac{1}{w^+_k} +
\displaystyle\sum_{k=0}^{M-1} \,\frac{1}{w^+_k} \,
\sum_{i=k+1}^{M} \prod_{j=k+1}^{i} \;\frac{w^-_j}{w^+_j} \;,
\\
T(1) &=& T(0) - \displaystyle\frac{1}{w^+_0} \;,
\\
T(n) &=& T(0) -
\displaystyle\sum_{k=0}^{n-1} \,\displaystyle\frac{1}{w^+_k} -
\displaystyle\sum_{k=0}^{n-2} \,\displaystyle\frac{1}{w^+_k} \,
\displaystyle\sum_{i=k+1}^{n-1} \prod_{j=k+1}^{i} \;\frac{w^-_j}{w^+_j}
\;\; (2 \leq n \leq M) \,.
\label{MFPT:ra}
\end{array}
\end{equation}
In our model with $M$ servers, using Eq.~(\ref{Mservers})
and the parameter $\gamma$ defined in the text,
we can recast the products in Eq.~(\ref{MFPT:ra}) as
\begin{equation}
\prod_{j=k+1}^{i} \;\frac{w^-_j}{w^+_j} =
\gamma^{i-k} \prod_{j=k+1}^{i} j = \gamma^{i-k} \,\frac{i!}{k!} \,.
\label{prod1}
\end{equation}
Thus, we can also recast the sums in Eq.~(\ref{MFPT:ra}) as
\begin{equation}
\sum_{i=k+1}^{n-1} \prod_{j=k+1}^{i} \;\frac{w^-_j}{w^+_j} =
\frac{\gamma^{-k}}{k!} \sum_{i=k+1}^{n-1} i! \,\gamma^i \,,
\label{Sprod1}
\end{equation}
and
\begin{equation}
\sum_{k=0}^{n-2} \,\frac{1}{w^+_k} \,
\sum_{i=k+1}^{n-1} \prod_{j=k+1}^{i} \;\frac{w^-_j}{w^+_j} =
\frac{1}{\lambda} \sum_{k=0}^{n-2} \,
\frac{\gamma^{-k}}{k!} \sum_{i=k+1}^{n-1} i! \,\gamma^i \,.
\label{SSprod1}
\end{equation}
Replacing the last two expressions in Eq.~(\ref{MFPT:ra}),
we obtain Eq.~(\ref{MFPT:critical}) in Sec.~\ref{Sec:Critical}
in the main text.

%%%%%%%%%%%%%%%%%%%%%%%%%%%%%%%%%%%%%%%%%%%%%%%%%%%%%%%%%%%%%%%%%%%%%%
\subsection{Steady-state}
\label{sec:SS}
%%%%%%%%%%%%%%%%%%%%%%%%%%%%%%%%%%%%%%%%%%%%%%%%%%%%%%%%%%%%%%%%%%%%%%

Following~\cite{Win03}, we can construct the steady-state
of the problem. From Eq.~(\ref{mastereq}), the time independent
solution must satisfy
\begin{equation}
\begin{array}{rcl}
\omega^-_1 \,\pi_1 -\omega^+_0 \,\pi_0 &=& 0 \,,
\\
\rule{0cm}{0.5cm}
\omega^-_{n+1} \,\pi_{n+1} +\omega^+_{n-1} \,\pi_{n-1}
- (\omega^+_n + \omega^-_n) \,\pi_n &=& 0
\;\; \mbox{$(n \geq 1)$} \,.
\end{array}
\label{stationary}
\end{equation}
Thus, from the first expression we immediately obtain
\begin{equation}
\pi_1 = \displaystyle\frac{\omega^+_0}{\omega^-_1} \,\pi_0 \,.
\label{P1}
\end{equation}
Substituting this result in the second equation ($n=1$) yields
\begin{equation}
\pi_2 = \displaystyle
\frac{\omega^+_1 \omega^+_0}{\omega^-_2 \omega^-_1} \,\pi_0 \,,
\label{P2}
\end{equation}
and so on we can proof by induction that
\begin{equation}
\pi_n = \frac{\omega^+_{n-1} \dots \omega^+_0}
           {\omega^-_n \dots \omega^-_1} \,\pi_0
= \prod_{j=1}^n \,\frac{\omega^+_{j-1}}{\omega^-_j} \,\pi_0\;
(n \geq 1)  \,.
\label{Pn}
\end{equation}
From the normalization condition, $\sum_{n=0}^{\infty} \pi_n = 1$,
results
\begin{equation}
\pi_0 = \frac{1}{S} \,,
\label{P0}
\end{equation}
where
\begin{equation}
S = 1 + \sum_{n=1}^{\infty}
\,\prod_{j=1}^n \,\frac{\omega^+_{j-1}}{\omega^-_j} \,.
\label{Sdef}
\end{equation}
The existence of the steady-state is determined by the convergence
of the series in the Eq.~(\ref{Sdef}).

For the model given by Eq.~(\ref{Mservers}),
the product in Eq.~(\ref{Sdef}) can be written as,
\begin{equation}
\prod_{j=1}^n \frac{\omega^+_{j-1}}{\omega^-_j}
= \left\{
\begin{array}{ll}
\displaystyle
\left( \frac{\lambda}{\mu} \right)^n \;\prod_{j=1}^n \,\frac{1}{j}
= \left( \frac{\lambda}{\mu} \right)^n \,\frac{1}{n!}
& (0 \leq n \leq M) \,,
\\
\rule{0cm}{1.0cm}
\displaystyle
\left( \frac{\lambda}{\mu} \right)^n
\;\prod_{j=1}^{M-1} \,\frac{1}{j} \,\prod_{j=M}^{n} \,\frac{1}{M}
= \left( \frac{\lambda}{\mu} \right)^n \,\frac{1}{(M-1)!}
\,\frac{1}{M^{n-(M-1)}}
& (n \geq M) \,.
\end{array}
\right.
\label{prod}
\end{equation}
Substituting Ec.~(\ref{prod}) into Eq.~(\ref{Sdef}) we obtain,
\begin{equation}
\begin{array}{lcl}
S &=& \displaystyle
1 + \sum_{n=1}^{M-1} \left( \frac{\lambda}{\mu} \right)^n
 \,\frac{1}{n!}
+ \sum_{n=M}^{\infty} \left( \frac{\lambda}{\mu} \right)^n
 \,\frac{1}{(M-1)!} \,\frac{1}{M^{n-(M-1)}} \,,
\\
\rule{0cm}{1.0cm}
&=& \displaystyle
\sum_{n=0}^{M-1} \,\frac{(M \,\rho)^n}{n!}
+ \frac{M^{M-1}}{(M-1)!} \sum_{n=M}^{\infty} \,\rho^n \,,
\end{array}
\label{S1}
\end{equation}
where $\rho$ is given by Eq.~(\ref{rho}).
The convergence of series in the last expression only occurs
for $\rho<1$. In this case,
\begin{equation}
\sum_{n=M}^{\infty} \rho^n = \sum_{i=0}^{\infty} \rho^{i+M}
= \rho^M \,\sum_{i=0}^{\infty} \rho^i = \frac{\rho^M}{1 - \rho} \,,
\label{seriesg}
\end{equation}
and from Eqs.~(\ref{Pn}),~(\ref{prod}),
and~(\ref{S1}),~(\ref{seriesg}), we obtain
the Eqs.~(\ref{PnM}) and~(\ref{S})
in the main text, respectively.

%%%%%%%%%%%%%%%%%%%%%%%%%%%%%%%%%%%%%%%%%%%%%%%%%%%%%%%%%%%%%%%%%%%%%%
%%%%%%%%%%%%%%%%%%         Acknowledgments         %%%%%%%%%%%%%%%%%%%
%%%%%%%%%%%%%%%%%%%%%%%%%%%%%%%%%%%%%%%%%%%%%%%%%%%%%%%%%%%%%%%%%%%%%%

\vspace{0.2in}
\subsection*{Acknowledgment}
An early stage of this work was partially supported by
{\em Sistema de Urgencias del Rosafe SA}, C{\'o}rdoba, Argentina.

%%%%%%%%%%%%%%%%%%%%%%%%%%%%%%%%%%%%%%%%%%%%%%%%%%%%%%%%%%%%%%%%%%%%%%
%%%%%%%%%%%%%%%%%%            References           %%%%%%%%%%%%%%%%%%%
%%%%%%%%%%%%%%%%%%%%%%%%%%%%%%%%%%%%%%%%%%%%%%%%%%%%%%%%%%%%%%%%%%%%%%

\vspace{0.2in}
%\bibliographystyle{unsrt}
%\bibliography{queues}

%%%%%%%%%%%%%%%%%%%%%%%%%%%%%%%%%%%%%%%%%%%%%%%%%%%%%%%%%%%%%%%%%%%%%%
\end{document}